\documentclass[useAMS,usenatbib]{mn2e}
\usepackage{amscd}
\usepackage{txfonts}
\usepackage{graphicx}
\usepackage{natbib}

\def\del#1{{}}

\newcommand{\bra}{\langle}
\newcommand{\ket}{\rangle}
\newcommand{\dd}{\mathrm{d}}
\newcommand{\ltsima}{$\; \buildrel < \over \sim \;$}
\newcommand{\lsim}{\lower.5ex\hbox{\ltsima}}
\newcommand{\gtsima}{$\; \buildrel > \over \sim \;$}
\newcommand{\gsim}{\lower.5ex\hbox{\gtsima}}

\title[Coded mask imaging of extended sources with Gaussian random fields]
{Coded mask imaging of extended sources \newline with Gaussian random fields}

\author[B. M. Sch\"afer]
{B. M. Sch\"afer\thanks{e-mail: spirou@mpa-garching.mpg.de}\\
Max-Planck-Institut f\"ur Astrophysik, Karl-Schwarzschild-Stra{\ss}e 1, Postfach 1317, 85741 Garching, Germany}

\begin{document}
\pagerange{\pageref{firstpage}--\pageref{lastpage}}
\pubyear{2003}
\maketitle
\label{firstpage}

\begin{abstract}
A novel method for generating coded mask patterns based on Gaussian random fields (GRF) is proposed. In contrast to
traditional algorithms based on cyclic difference sets, it is possible to construct mask patterns that encode
a predefined point spread function (PSF). The viability of this approach and the reproducibility of the PSFs
is examined, together with studies on the mean transparency, pixel-to-pixel variance and PSF deterioration due to
partial shadowing. Sensitivity considerations suggest the construction of thresholded realisations of Gaussian random 
fields (TGRF) which were subjected to the same analyses. Special emphasis is given to ray-tracing simulations of the pattern's 
performance under finite photon statistics in the observation of point sources as well as of extended sources in comparison to 
random masks and the pattern employed in the wide field imager onboard {\em BeppoSAX}. A key result is that in contrast to 
traditional mask generation schemes, coded masks based on GRFs are able to identify extended sources at accessible photon 
statistics. Apart from simulating on-axis observations with varying levels of signal and background photon counts, partial 
shadowing of the mask pattern in the case of off-axis observations and the corresponding field-of-view is 
assessed.
\end{abstract}


\section{Introduction}
In X-ray astronomy, focusing of radiation is so far feasible only for photon energies up to about 10~keV 
through grazing incidence reflection. Applied in Wolter-type mirrors, this method can provide a very good angular resolution, 
i.e. down to $0\farcs5$ in the case of {\em Chandra}\footnote{\tt http://cxc.harvard.edu/} and $4\arcsec-12\arcsec$ for {\em 
XMM-Newton}\footnote{\tt http://xmm.vilspa.esa.es/}. The collecting area is maximised through the use of nested mirrors. The 
field-of-view (FOV) is limited by the grazing incidence condition set by the diffractive index of the mirror material to $\lsim 
1\degr$. At energies higher than 10 keV, focusing is technologically very hard to archieve. A workaround are coded mask 
imagers, where a position sensitive detector records the shadow of a mask pattern cast by the sources under investigation. The 
arrangement of sources can be reconstructed by cross-correlating the recorded shadowgram with the mask pattern.

Coded masks have by now found a widespread use in high energy astrophysics and there is a large number of successful missions 
such as {\em BeppoSAX}\footnote{\tt http://bepposax.gsfc.nasa.gov/bepposax/index.html}, currently flying intruments like
{\em INTEGRAL}\footnote{\tt http://astro.estec.esa.nl/SA-general/Projects/Integral/} and {\em HETE-2}\footnote{\tt
http://space.mit.edu/HETE/}, and ambitious future projects, for instance {\em SWIFT}\footnote{\tt
http://swift.gsfc.nasa.gov/}.

In this paper, I propose coded mask patterns based on Gaussian random fields, because they enable the construction of a coded 
mask device for predefined imaging characteristics, i.e. for a given PSF. The shape of the PSF can be tuned to match the 
anticipated source profile. A beautiful example of a naturally occurring Gaussian random field is the pattern of fluctuations 
in the cosmic microwave background (CMB). Analyses of WMAP data carried out among others by \citet{2001MNRAS.326.1243C} and 
\citet{2003astro.ph..2223K} find the CMB consistent with Gaussian primordial fluctuations and have set upper limits on 
non-Gaussianity.

After a recapitulation of coded mask imaging and existing mask pattern generation schemes in Sect.~\ref{coded_mask_imaging}, 
GRFs are introduced in Sect.~\ref{gaussian_random_fields}. The feasibility of GRFs in coded 
mask imagers is examined in Sect.~\ref{results} with special emphasis on  the performance of GRFs in realistic scenarios, i.e. 
under finite photon statistics  and in the observation of extended sources (Sect.~\ref{finite_photon}). A summary of the key  
results in Sect.~\ref{summary} concludes the paper.

\section{Coded mask imaging}\label{coded_mask_imaging}
Coded mask cameras observe a source by recording the shadow cast by the mask onto the detector. The mask pattern is
described by the position dependent transparency $\psi(x)$. A shifted shadowgram $\psi(x-d\tan(\theta))$ is observed if
the radiation incides under an angle $\theta$ with respect to the optical axis. The distance between the coded mask and
the detector is denoted by $d$. The correlation function $a(x)$, defined as
\begin{equation}a(x) =
\psi(x-d\tan\theta)\otimes\psi(x)=\bra\psi(x-d\tan\theta+\lambda)\psi(\lambda)\ket_\lambda\mbox{,}
\end{equation}
peaks at $x_0 = d\tan\theta$, from which the angle of incidence $\theta = \arctan(x_0/d)$ can be inferred. The PSF
$c(x)$, defined as the correlation function at normal incidence ($\theta=0\degr$), i.e. the auto-correlation function, reads
\begin{equation}
c(x) = \psi(x)\otimes\psi(x)=\bra\psi(x+\lambda)\psi(\lambda)\ket_\lambda\mbox{.}
\end{equation}
The influence of imperfections of the detector can be modelled by convolution of $\psi(\bmath{x})$ with suitable kernels
describing the positional detector response \citep[see, e.g.][]{2003astro.ph..1132S}. Techniques for analysing
coded mask data have been summarised by \citet{1987Ap&SS.136..337S} and \citet{1987SSRv...45..349C}

Random mask patterns as used in the HETE-2 satellite \citep{1994STIN...9522436I} consist of white noise. They are not ideal 
imagers, because their auto-correlation possess sidelobes and are not perfectly flat. Aiming at $\delta$-like PSFs, mask 
patterns based on cyclic difference sets have been introduced by \cite{1976MNRAS.177..485G}. As pointed out by 
\cite{1978ApOpt..17..337F}, these uniformly redundant arrays (URA) provide even sampling at all spatial scales. URA patterns 
are less susceptible to noise compared to truly random arrays and their auto-correlation function is a $\delta$-spike with 
perfectly flat sidelobes in case of complete imaging. In this paper, I propose a method for constructing coded mask pattern 
encoding arbitrary PSFs. While the traditional masks are optimised for the observation of point sources, the PSFs of masks 
based on GRFs can be adjusted to the source profile of extended sources and make the observation of 
extended sources such as extended structures in the Milky Way possible.

\section{Gaussian random fields}\label{gaussian_random_fields}

\subsection{Definitions}
The statistical properties of a GRF are homogeneous and isotropic and the phases of different Fourier
modes are mutually uncorrelated and random. A consequence of the central limit theorem is then that the amplitudes
follow a Gaussian distribution. Due to all correlations above the two-point level being either vanishing in the case of
odd moments or being expressible in terms of two-point functions for even moments, the statistics of amplitude
fluctuations in a GRF is completely described by its power spectrum $P(k)$ (see eqn.~(\ref{eqn_powerspec})).

Because the imaging characteristics of coded mask imagers are described by the PSF, which is defined to be the auto-correlation 
function of their mask pattern, i.e. by their power spectrum in case of isotropic PSFs, GRFs provide a tool 
for generating mask patterns with predefined imaging characteristics.The theory of structure formation in cosmology and the 
description of the cosmic microwave background makes extensive use of GRFs 
\citep[c.f.][]{1999coph.book.....P,1998gafo.conf.....L}. Their application is commonplace in generating initial conditions for 
simulations of cosmic structure formation and in constructing mock CMB fields for simulating sub-millimetric observations.

\subsection{Algorithm}
\label{algorithm}
Starting from the PSF $c(\bmath{x})$, the Fourier transform $C(\bmath{k})$ is derived:
\begin{equation}
C(\bmath{k}) = \mathcal{F}\left[c(\bmath{x})\right] =
\int\frac{\dd^2x}{(2\pi)^2}\:c(\bmath{x})\exp(-\mathrm{i}\bmath{kx})\mbox{.}
\end{equation}
The power spectrum $P(k)$ is defined as the Fourier-transform $C(\bmath{k})$ of the auto-correlation function 
$c(\bmath{x})$. In more than one dimension, an average of the Fourier transform $C(\bmath{k})$ of the statistically isotropic 
random field $c(\bmath{x})$ over all directions of the wave vector $\bmath{k}$ at fixed length $k=\left|\bmath{k}\right|$ needs 
to be performed:
\begin{equation}
P(k) = \bra\left| C(\bmath{k})\right|\ket_{\left|\bmath{k}\right|=k}\mbox{.}
\label{eqn_powerspec}
\end{equation}

All elementary waves $\exp(\mathrm{i}\bmath{kx})$ with wave vectors in the $k$-space shell
$\left[\left|\bmath{k}\right|,\left|\bmath{k}+\Delta\bmath{k}\right|\right]$ contribute to the variance $\sigma_{k}^2
=P(k)$ required by the power spectrum on scale $k=\left|\bmath{k}\right|$. In discretising, the amplitudes $\Psi(\bmath{k})$ 
are set such that their quadratic sum
$\sum_{k\in\left[\left|\bmath{k}\right|,\left|\bmath{k}+\Delta\bmath{k}\right|\right]}\left|\Psi(\bmath{k})\right|^2$
matches $\sigma_{k}^2$ with the only exception $\Psi(\bmath{k} = 0)$, which is set to zero in order to ensure a
vanishing expectation value of the realisation $\psi(\bmath{x})$. The normal modes $\exp(\mathrm{i}\bmath{kx})$ are modified by 
a phase factor $\exp(2\pi\mathrm{i}q)$, where $q\in\left[0,1\right)$ is a uniformly distributed random number. By inverse 
Fourier transform, the normal modes $\Psi(\bmath{k})$ are brought to interference which finally results in the realisation, the 
real part of which is denoted by $\psi(\bmath{x})$:
\begin{equation}
\psi(\bmath{x})=\Re\left(\mathcal{F}^{-1}\left[\Psi(\bmath{k})\right]\right)=
\Re\left(\int\dd^2k\:\Psi(\bmath{k})\exp(\mathrm{i}\bmath{kx}+2\pi\mathrm{i}q)\right)\mbox{.}
\end{equation}

Alternatively, one may require the additional symmetry $\Psi(-\bmath{k})= \Psi^{*}(\bmath{k})$ in Fourier space (the complex 
conjugation is denoted by the asterisk), which forces the realisation to be purely real. The flow chart eqn.~(\ref{diagram}) 
summarises all steps:

\begin{center}
\begin{equation}
\begin{CD}C(\bmath{k})@>{\bra\ldots\ket_{\left|\bmath{k}\right| = k}}>>
P(k)@>{\cdot\exp(2\pi\mathrm{i}q)}>>\Psi(\bmath{k})			\\@A{\mathcal{F}}AA@.@VV{\mathcal{F}^{-1}}V
\\c(\bmath{x})@.@.\psi(\bmath{x})\mbox{.}
\end{CD}
\label{diagram}
\end{equation}
\end{center}

Due to the periodic boundary conditions imposed by the Fourier transform, the resulting realisations of the Gaussian
random field have cyclic boundaries, which is a desirable feature for coded mask patterns. For reasons of numerical
accuracy, it is strongly recommended to use shells in $\bmath{k}$-space with varying thickness
$\Delta\bmath{k}\propto\left|\bmath{k}\right|^{-1}$, such that approximately the same number of discretised
modes contributes to the variance required by the power spectrum $P(k)$.

\subsection{Choice of the PSF}
Although the algorithm outlined in Sect.~\ref{algorithm} is capable of generating random fields
$\psi(\bmath{x})$ encoding any isotropic PSF $c(\bmath{x})$, PSFs should be shaped like Lorenzian functions $c_L(x)$ or 
Gaussian functions $c_G(x)$. The parameter $\sigma_x$ describes the spatial extent:

\begin{eqnarray}
c_L(x) & = & \frac{\sigma_x^2}{x^2 + \sigma_x^2}\mbox{,}\\
c_G(x) & = & \exp\left(-\frac{x^2}{2\sigma_x^2}\right)\mbox{.}
\end{eqnarray}

The normalisation has been chosen such that the maximum correlation strength at $x=0$ is set to one. In the realisation 
$\psi(\bmath{x})$, the variable $\sigma_x$, that parameterises the PSF can be interpreted as a correlation length. 
\cite{1993A&A...276..673S} have pursued a related idea and have suggested coded masks with two spatial scales. In contrast, the 
realisations considered here have an entire spectrum of length scales.

\subsection{Scaling applied to the Gaussian random fields}
If one aims at employing GRFs in coded mask imagers, the field has to be scaled such that it assumes
values ranging from $\psi(\bmath{x})=0$ (opaqueness) to full transparency ($\psi(\bmath{x})=1$). This 
scaling ensures that the full dynamical range between is used and the modulation of the shadowgram as strong as 
possible. Hence, the sensitivity is maximised. One could think of two different linear transformations, the most 
intuitive being:

\begin{equation}
\psi(\bmath{x})\longrightarrow\psi^\star(\bmath{x}) =
\frac{\psi(\bmath{x})-\mathrm{min}\left\{\psi(\bmath{x})\right\}}
{\mathrm{max}\left\{\psi(\bmath{x})\right\} - \mathrm{min}\left\{\psi(\bmath{x})\right\}}\mbox{.}
\label{scaling1}
\end{equation}

With the symmetry condition $\mathrm{max}\left\{\psi(\bmath{x})\right\} = -\mathrm{min}\left\{\psi(\bmath{x})\right\}$
being fulfilled, the mean transparency $\bra\psi^\star\left(\bmath{x}\right)\ket$ is equal to 1/2: The mean
$\bra\psi\left(\bmath{x}\right)\ket = 0$ vanishes by construction, because each normal mode $\cos(\bmath{kx})$ has a
vanishing expectation value. In general, the realisation $\psi(\bmath{x})$ will not fulfill the above mentioned symmetry
condition. 

Instead, the scaling
\begin{equation}
\psi(\bmath{x})\longrightarrow\psi^\prime(\bmath{x}) =
\frac{1}{2}
\left[\frac{\psi(\bmath{x})}{\mathrm{max}\left\{\left|\psi(\bmath{x})\right|\right\}} + 1\right]\mbox{}
\label{scaling2}
\end{equation}
ensures $\bra\psi^\prime\ket=1/2$ and will be used in the remainder of the paper. It should be noted that none of the
above scalings strictly conserves Gaussianity, because each particular realisation is scaled by its maximal amplitude and 
consequently, high amplitudes do not appear any more in an ensemble of realisations.

Now that the mean transparency $\bra\psi^\prime\ket$ is fixed, the absolute flux from a source can be inferred from the
number of measured photons. The scaling eqn.~(\ref{scaling1}) may be taken advantage of in designing a mask that
blocks a larger or smaller fraction of photons than the generic fraction of 1/2: In anticipation of 
Sect.~\ref{pixel2pixelvariance}, in the case of a realisation of a GRF encoding a Gaussian PSF 
$c_G(x)$ with $\sigma_x = 8$ pixels, the probability density $p\left(\bra\psi^\star\ket\right)\dd\bra\psi^\star\ket$ of 
the mean transparency $t=\bra\psi^\star\ket$ is described by a Gaussian distribution with mean $\mu_t =0.504\pm0.082$ 
and standard deviation $\sigma_t =0.028\pm0.006$ at 95\% confidence. When constructing realisations of Gaussian fields 
for coded mask instruments, one obtains patterns with transparencies 
$\bra\psi^\star\ket\in\left[\mu_t-\sigma_t,\mu_t+\sigma_t\right]$ with a probability of 
$\mathrm{erf}(1/\sqrt{2})\simeq0.6827$.

\subsection{Gaussian random fields for circular apertures}
\begin{figure}
\begin{tabular}{cc}
\resizebox{3.7cm}{!}{\includegraphics{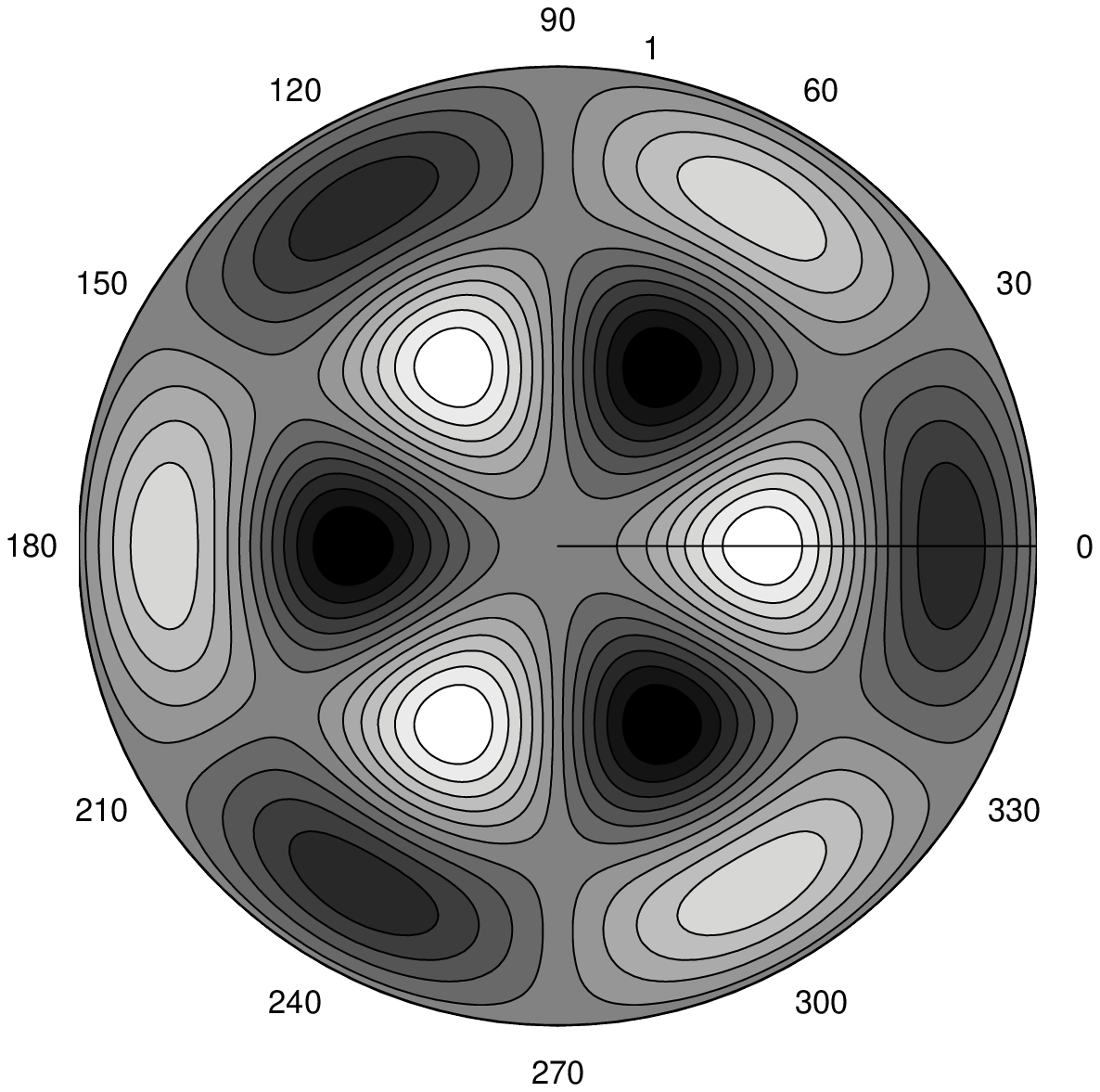}}
&
\resizebox{3.7cm}{!}{\includegraphics{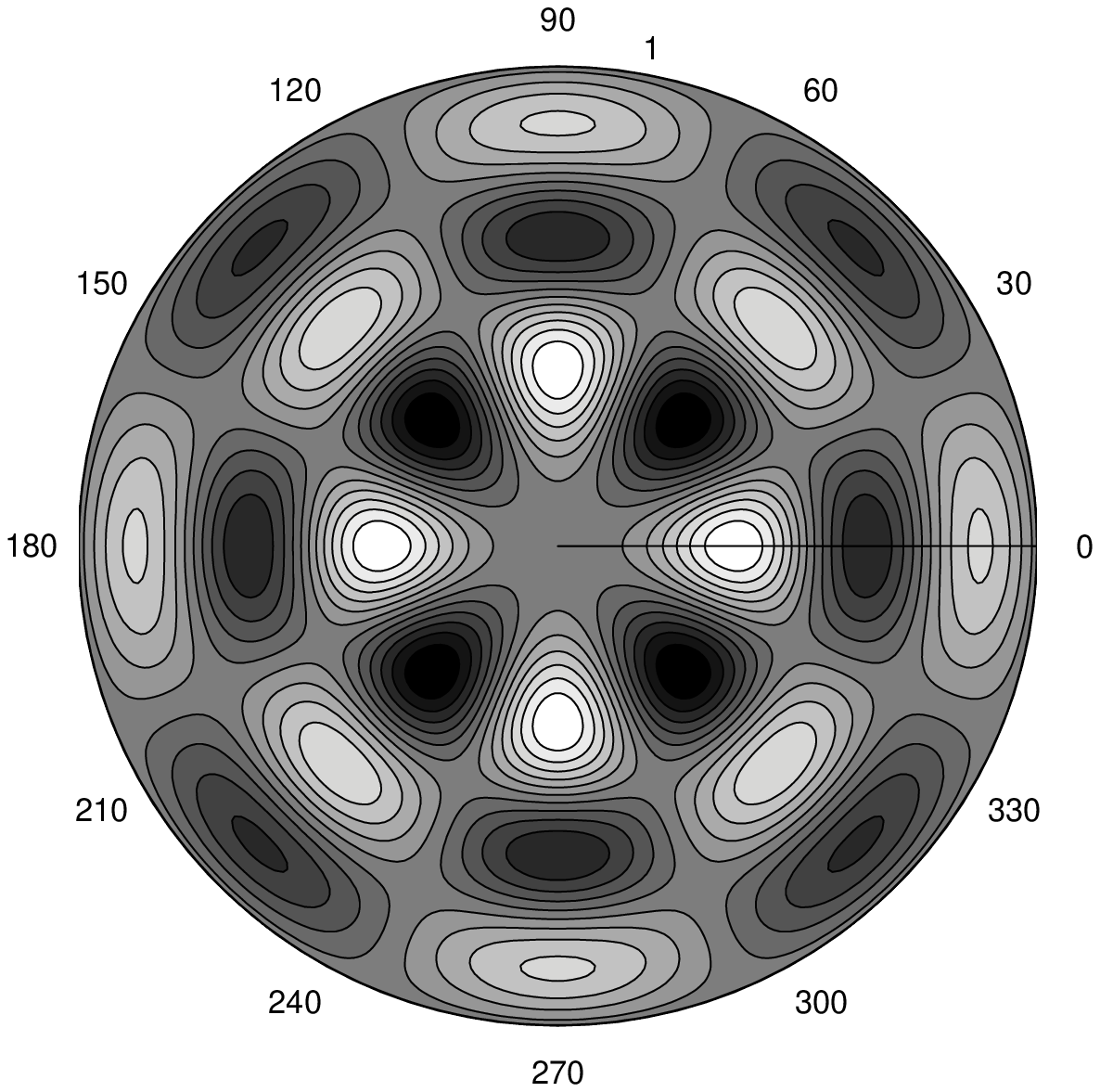}}
\end{tabular}
\caption{Normal modes $\psi_{\ell m}(r,\phi)$ used for constructing GRFs for circular apertures, for 
$(\ell,m)=(2,3)$ (left panel) and $(\ell,m)=(3,4)$ (right panel).}
\label{fig_bessel}
\end{figure}

For coded-mask experiments with a circular aperture it is possible to construct GRFs with azimuthal 
symmetry, in the same way as hexagonal uniformaly redundant arrays (HURA) are an adaptation of the URA patterns to circular 
apertures \citep{1985ICRC....3..295F}. Instead of constructing a GRF with plane waves as the solutions of Laplace's equation
$\triangle\psi(\bmath{x})=\left(\partial_x^2+\partial_y^2\right)\psi(\bmath{x})=0$ in Cartesian coordinates 
($\bmath{x}=(x,y)$) with boudnary conditions $\psi(\bmath{x}=-L)=0=\psi(\bmath{x}={L})$ ($2L$ denotes the pattern's side 
length) one would resort to solving 
$\triangle\psi(\bmath{r})=\left(\partial^2_r+1/r\partial_r+1/r^2\partial^2_\phi\right)\psi(\bmath{r})=0$ in polar coordinates 
($\bmath{r}=(r,\phi)$) with the boundary condition $\psi(r=R)=0$ $\forall\phi$, where the radius of the aperture is denoted as 
$R$. $\psi$ is easily found as the solution to Bessel's differential equation and reads as:
\begin{equation}
\psi_{\ell m}(r,\phi) = J_m\left[r\cdot Z_m(\ell)\right]\cdot\exp(i m\phi)\mbox{,}
\label{eqn_bessel_basis}
\end{equation}
where the numbers $\ell$ and $m$ are only allowed to assume integer values. $Z_m(\ell)$ is the $\ell^\mathrm{th}$ zero of the 
Bessel function $J_m$. In Fig.~\ref{fig_bessel}, two solutions are depicted for $(\ell,m)=(2,3)$ and $(\ell,m)=(3,4)$. In 
reality, it might be cumbersome to construct a GRF on the basis of the normal modes given by 
eqn.~(\ref{eqn_bessel_basis}) due to Bessel function's complicated orthonormality relations. 

\section{Results}\label{results}
In order to provide a visual impression, two GRFs encoding the above stated PSF with their auto-correlation functions 
are presented (Sect.~\ref{visual_impression}). Subsequently, the reproducibility of the chosen PSF 
(Sect.~\ref{reproducibility}), the pixel-to-pixel variance (Sect.~\ref{pixel2pixelvariance}), the Gaussianity of the 
distribution of pixel amplitudes (Sect.~\ref{pixamp_distro}) and the shape of the PSF under partial shadowing 
(Sect.~\ref{partial_shadowing}) are examined. Finally, thresholded GRFs are introduced and the deterioration 
of the PSF of such thresholded realisations (Sect.~\ref{sec_threshold}) is addressed.

\subsection{Visual impression}\label{visual_impression}
\begin{figure}
\begin{center}
\begin{tabular}{c}
\resizebox{7cm}{!}{\includegraphics{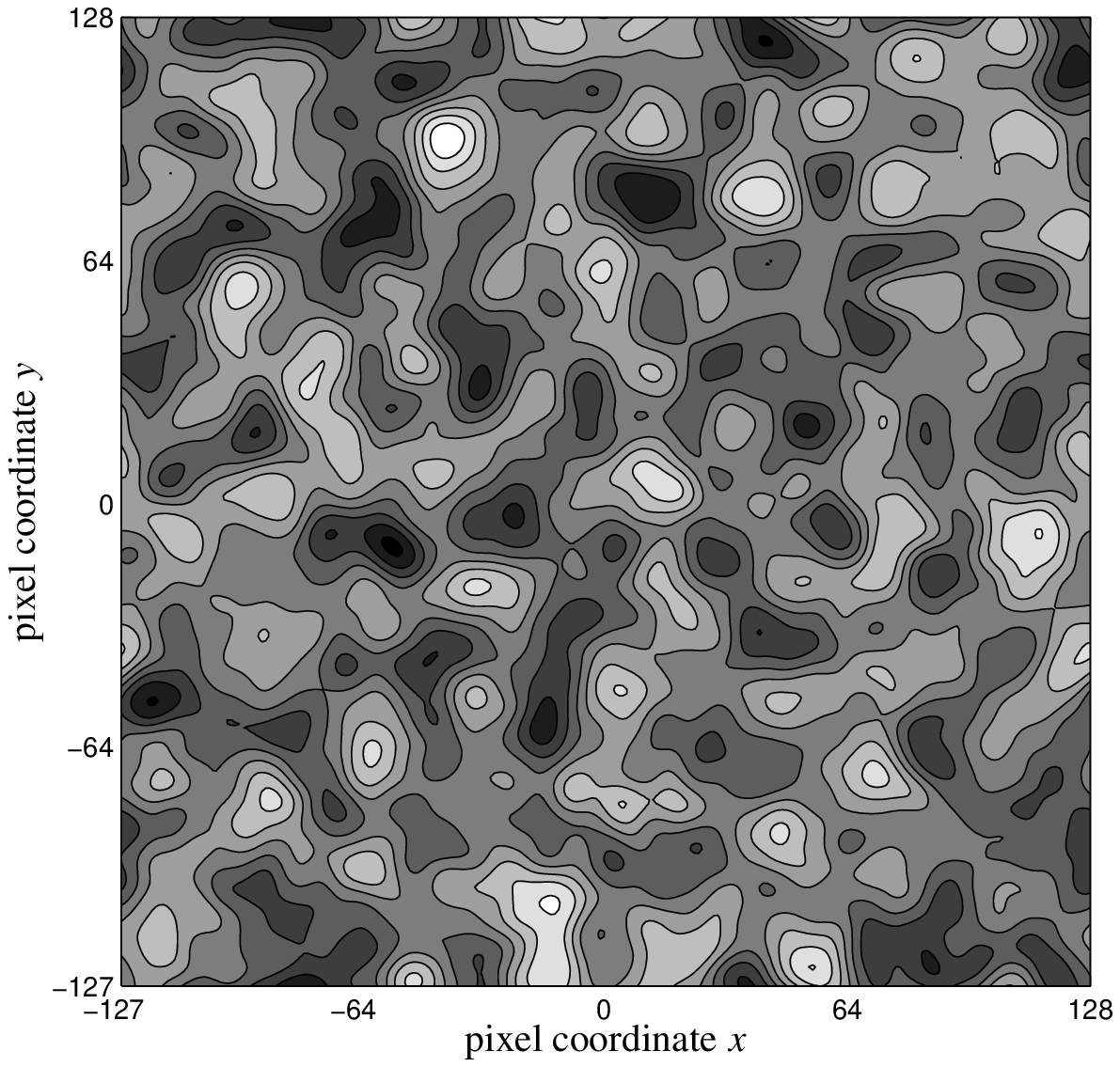}} \\
\resizebox{7cm}{!}{\includegraphics{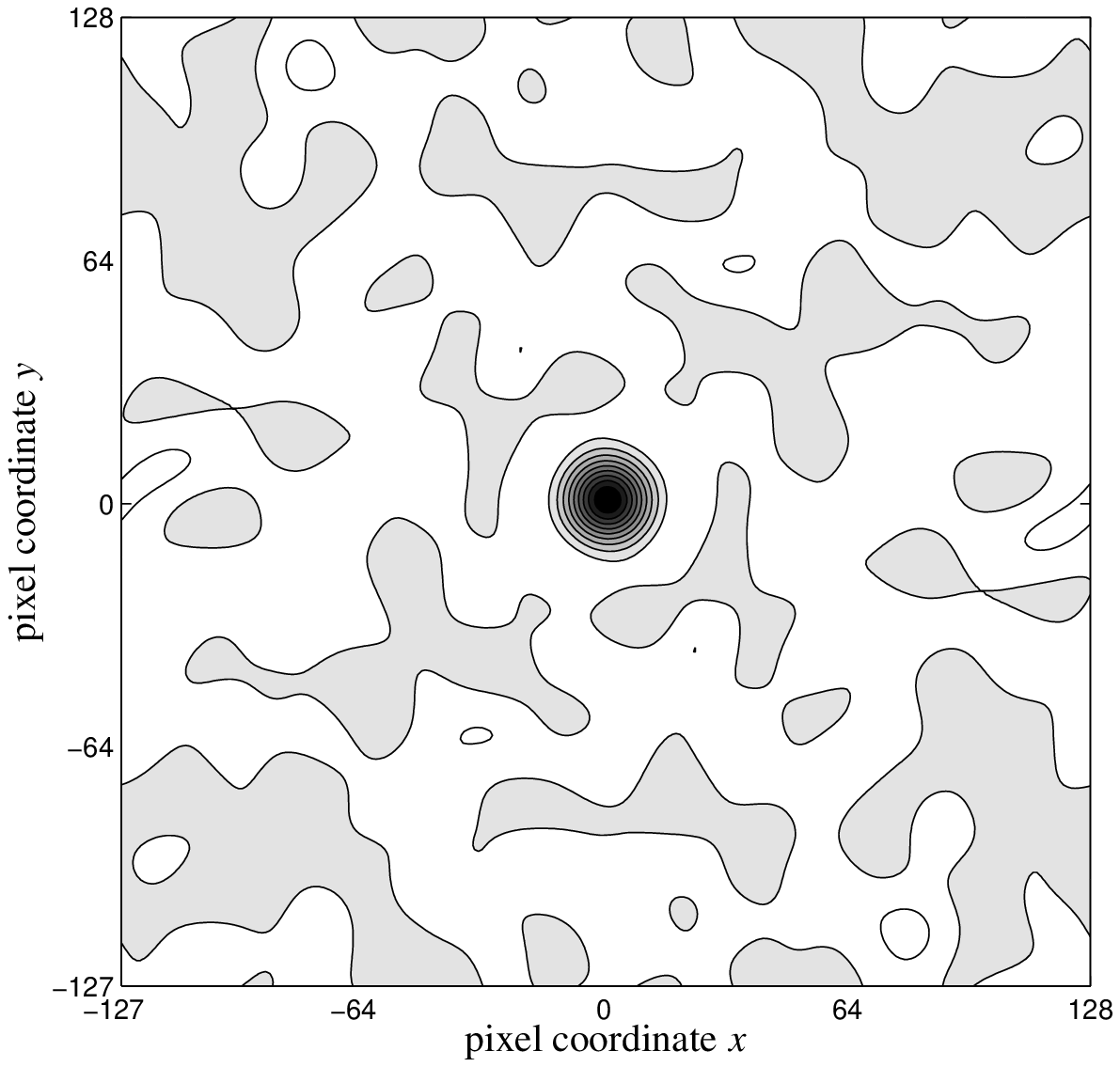}}
\end{tabular}
\end{center}
\caption{A realisation of a GRF $\psi_G(\bmath{x})$ (upper panel) for the Gaussian PSF $c_G(x)$
and the auto-correlation function $\psi_G(\bmath{x})\otimes\psi_G(\bmath{x})$ (lower panel).}
\label{fig_gauss_realise}
\end{figure}

\begin{figure}
\begin{center}
\begin{tabular}{c}
\resizebox{7cm}{!}{\includegraphics{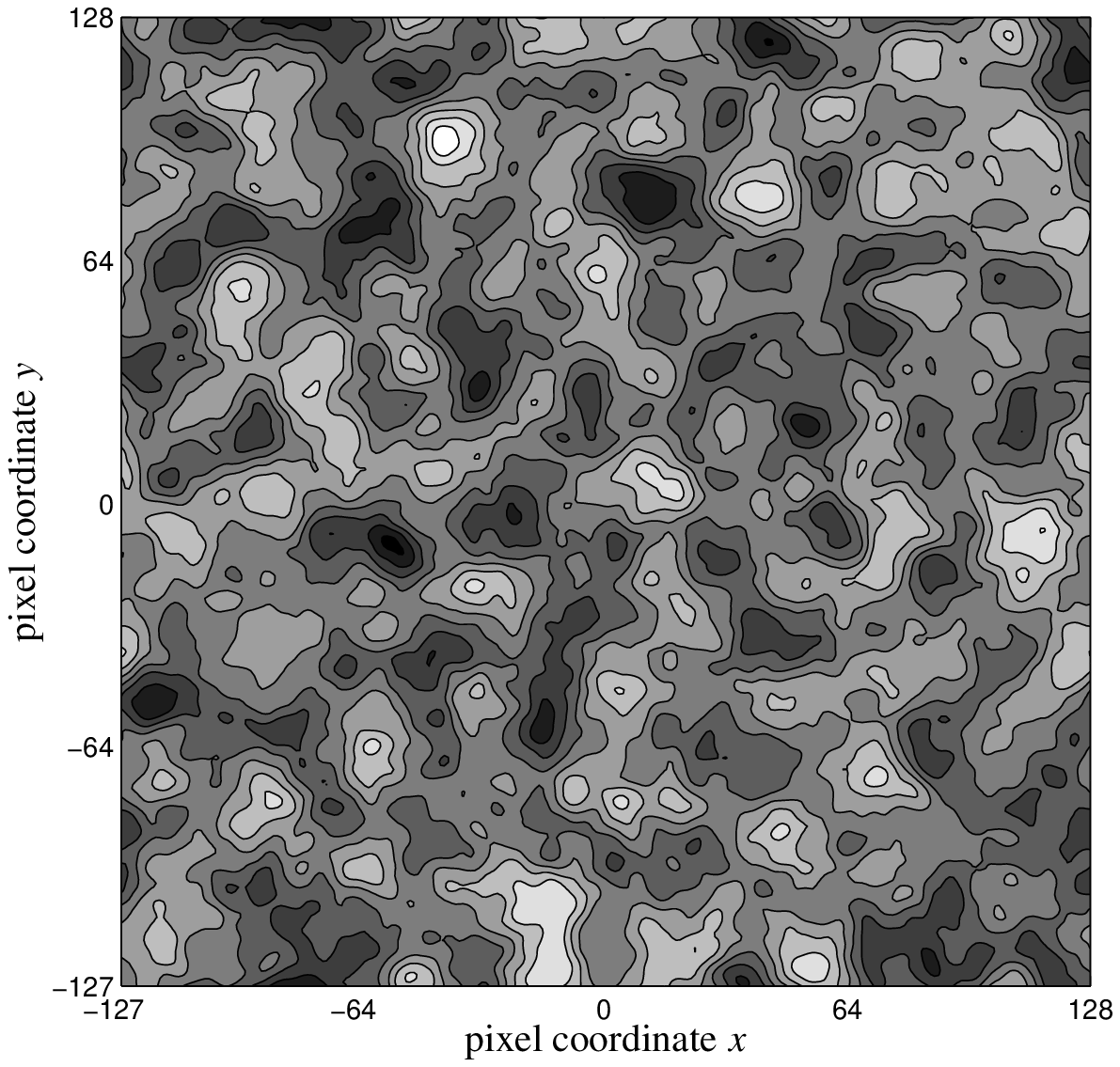}} \\
\resizebox{7cm}{!}{\includegraphics{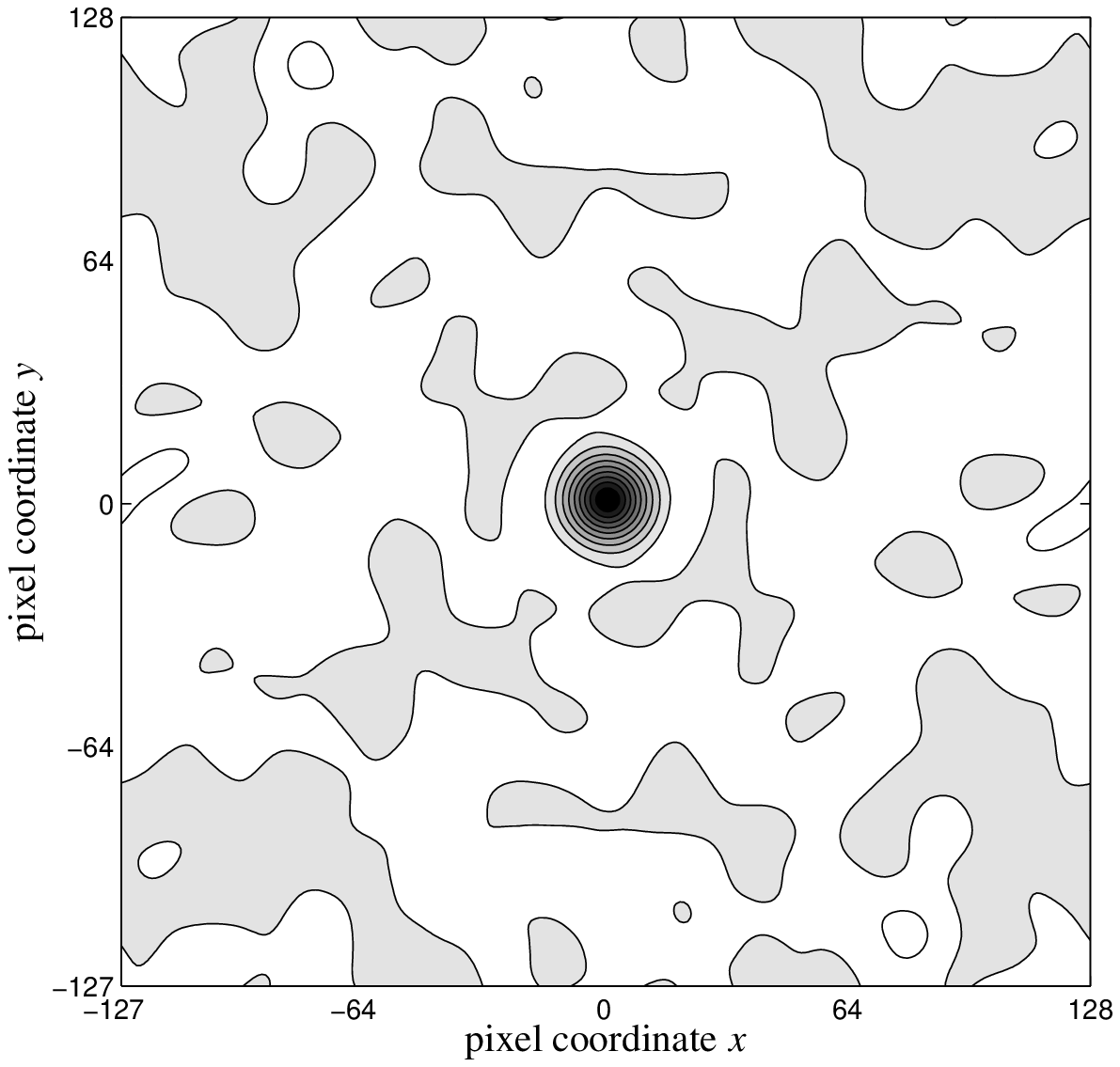}}
\end{tabular}
\end{center}
\caption{A realisation of a GRF $\psi_L(\bmath{x})$ (upper panel) for the Lorenzian PSF
$c_L(x)$ and the auto-correlation function $\psi_L(\bmath{x})\otimes\psi_L(\bmath{x})$ (lower panel).}
\label{fig_lorenz_realise}
\end{figure}

Following the above prescription, 100 realisations of GRFs encoding Gaussian and Lorenzian
PSFs of different widths $\sigma_x$ were generated on a 2-dimensional square grid with $256^2$ mesh cells.
Figs.~\ref{fig_gauss_realise} and \ref{fig_lorenz_realise} show a realisation of the GRF and
its auto-correlation function for a Gaussian and a Lorenzian PSF, respectively. In order to facilitate comparison,
the widths of the PSFs have been chosen to be the same: $\sigma_x = 8\mbox{ pixels}$. The random fields are
scaled to mean values of 1/2 (by means of eqn.~(\ref{scaling2})) and the central correlation strength in the
auto-correlation functions is equal to 1. The contours have a linear spacing of 0.1. The auto-correlation functions have
the symmetry property that $\psi(\bmath{x})\otimes\psi(\bmath{x})=\psi(-\bmath{x})\otimes\psi(-\bmath{x})$. In the derivation 
of auto-correlation and cross-correlation functions, the balanced correlation scheme was used. The correlation functions were 
derived for ideal detectors, i.e. finite position resolution or similar imperfections were neglected.

In comparing the realisations in Figs.~\ref{fig_gauss_realise} and \ref{fig_lorenz_realise} one notices the larger
abundance of small scale structures in the realisation encoding the Lorenzian PSF $c_L(x)$ in comparision to the
realisation derived for the Gaussain PSF $c_G(x)$. This can be explained by the fact that the power spectrum $P_L(k)$
declines $\propto\exp(-k)$ and thus much slower than the power spectrum $P_G(k)\propto\exp(-k^2)$. Both realisations
have been derived with the same random seed, i.e. the relative phases are identical and one immediately recognises
similar structures in $\psi_G(\bmath{x})$ and $\psi_L(\bmath{x})$.

\subsection{Reproducibility of the PSF}\label{reproducibility}
\begin{figure}
\resizebox{\hsize}{!}{\includegraphics{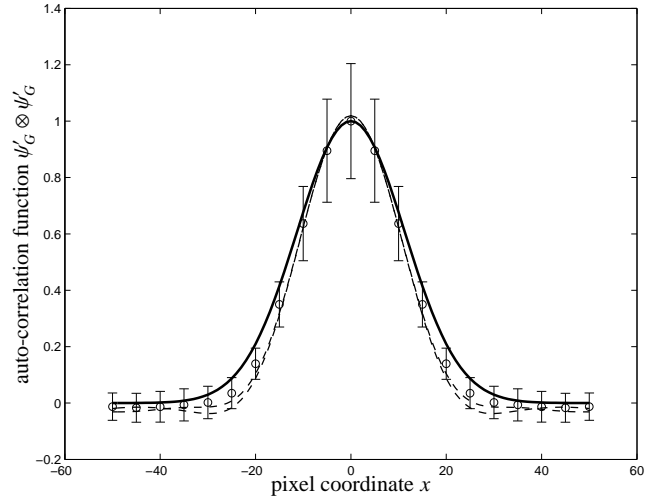}}
\caption{Cross section along the $x$-axis through the central part of the auto-correlation function
$\psi^\prime_G(\bmath{x})\otimes\psi^\prime_G(\bmath{x})$ for two different realisations $\psi^\prime_G(\bmath{x})$
(dashed) and the Gaussian target curve $c_G(x)$ (solid).}
\label{fig_gauss_psf}
\end{figure}

\begin{figure}
\resizebox{\hsize}{!}{\includegraphics{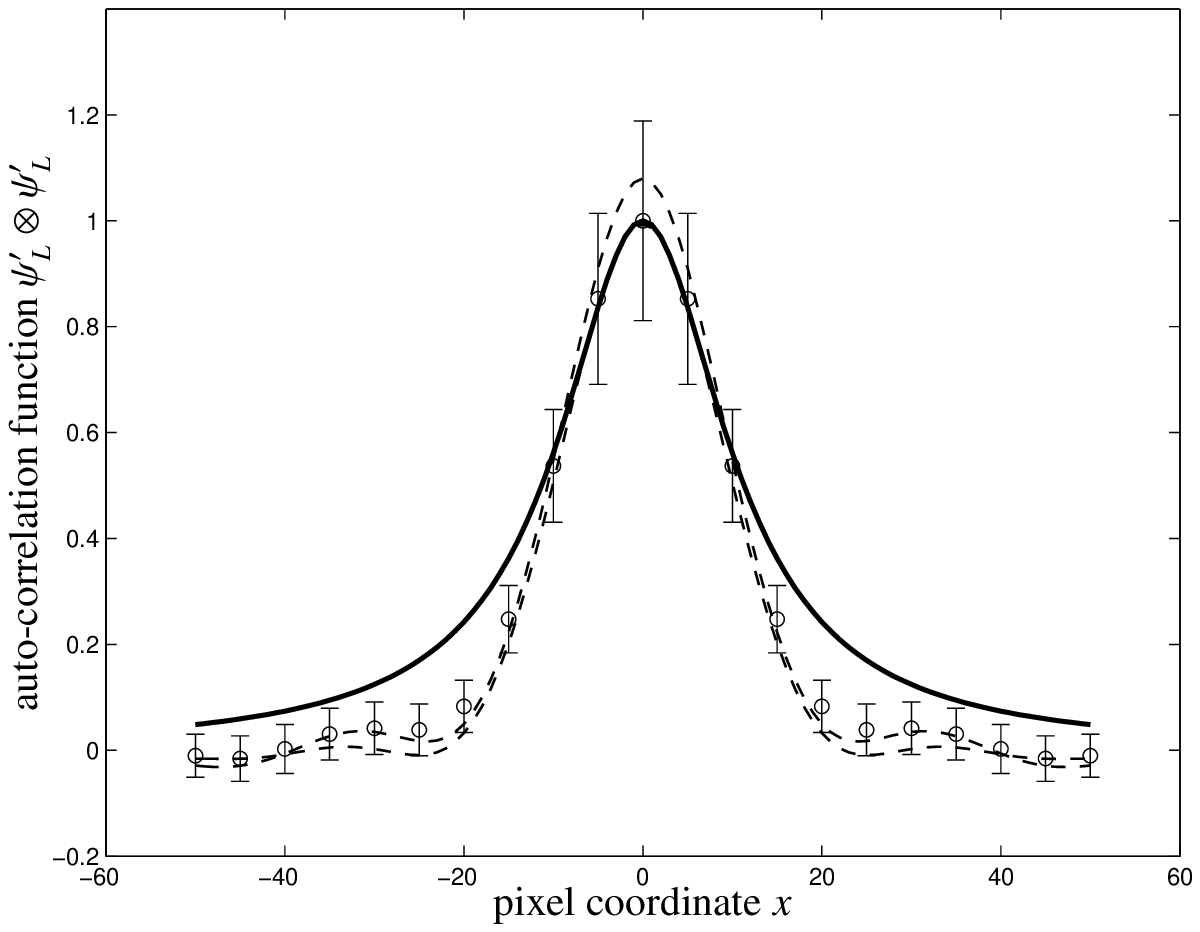}}
\caption{Cross section along the $x$-axis through the central part of the auto-correlation function
$\psi^\prime_L(\bmath{x})\otimes\psi^\prime_L(\bmath{x})$ for two different realisations $\psi^\prime_L(\bmath{x})$
(dashed) and the Lorenzian target curve $c_L(x)$ (solid).}
\label{fig_lorenz_psf}
\end{figure}

An important issue is the reproducibility of a chosen PSF $c(x)$ in realisations generated with differing random seeds.
This can be assessed by determining the auto-correlations of the scaled GRFs $\psi^\prime(\bmath{x})$ 
for all realisations within the data sample. In Fig.~\ref{fig_gauss_psf} the Gaussian target PSF $c_G(x)$ and the
auto-correlation functions $\psi^\prime_G(\bmath{x})\otimes\psi^\prime_G(\bmath{x})$ following from two realisations
$\psi(\bmath{x})$ are shown. The error bars denote the sample variance derived from 100 realisations of
$\psi_G(\bmath{x})$ following from different random seeds. The width of the PSF was chosen as $\sigma_x =
8\sqrt{2}\mbox{ pixels}$ for better visibility. Fig.~\ref{fig_lorenz_psf} shows the analogous for the Lorenzian target PSF 
$c_L(x)$ with $\sigma_x=8\sqrt{2}\mbox{ pixels}$.

As Figs.~\ref{fig_gauss_psf} and \ref{fig_lorenz_psf} illustrate, the functional shape of the target PSF $c(x)$ can be
reproduced with high reliability and the ratio of the peak-height to the correlation noise is $\simeq 40$. However,
there are minor imaging artefacts, namely very weak sidelobes: This is readily explained by the fact that the Fourier
transform of a well localised PSF in real space is extended and affected by the cutoff at the Nyquist frequency
$k_\mathrm{Nyquist}$, which induces a $\sin(k_\mathrm{Nyquist}x)/x$-like modulation. Consequently, the sidelobes are
suppressed in PSFs with large $\sigma_x$. The Lorenzian PSF is a bad choice in comparison to the Gaussian PSF, because
its Fourier transform $C(k)\propto\exp(-k)$ decays slower and is consequently more affected by the cutoff at
$k_\mathrm{Nyquist}$. Interpreting $\sigma_x$ as the correlation length of the GRF, it is clear that
in the limit of very narrow PSFs $\sigma_x$ assumes very small values, i.e. the amplitudes $\psi(\bmath{x})$ for
neighbouring pixels start loosing their correlation. This, however, does not correspond to white noise masks because the 
amplitude distribution is still Gaussian (c.f. Sect.~\ref{pixamp_distro}) and not bimodal, as in the case of white noise masks.

Due to the high confidence with which a chosen PSF is reproduced, the number of realisations to be examined is very
small. On the contrary, relying on truly random patterns, the number of necessary realisations with the accompanying
tests may be very high: For HETE-2, where such a random pattern is used, $10^5$ realisations had to be generated that
were subjected to certain boundary conditions \citep[see][]{1994STIN...9522436I}.

\subsection{Pixel-to-pixel variance}\label{pixel2pixelvariance}
\label{sec_variance}
In sensitivity considerations carried out by \cite{1994A&A...288..665I} for purely random masks, i.e. masks consisting
of either transparent ($\psi^\prime(\bmath{x})= 1$) or opaque ($\psi^\prime(\bmath{x}) = 0$) pixels, optimised mean
transparency $\bra\psi^\prime\ket$ and standard deviation $\sqrt{\bra\psi^{\prime 2}\ket - \bra\psi^\prime\ket^2}$ are
derived to be equal to 1/2. In that way, the variance and therefore the modulation of the signal is maximised. For the
GRFs considered here, the variance and hence the modulation of the shadowgram is noticably 
smaller. In Table~\ref{table_psistat}, the mean transparencies $\bra\psi^\prime\ket$, the variance
$\bra\psi^{\prime2}\ket-\bra\psi^\prime\ket^2$ and the standard deviation $\sqrt{\bra\psi^{\prime2}\ket
-\bra\psi^\prime\ket^2}$ together with their respective uncertainties for a set of GRFs encoding
Gaussian PSFs with differing width $\sigma_x$ are summarised.

\begin{table}
\vspace{-0.1cm}
\begin{center}
\begin{tabular}{cccc}
\hline\hline
\vphantom{\Large A}%
PSF width & mean transparency & variance & standard deviation \\
$\sigma_x$ & $\bra\psi^\prime\ket$ & $\bra\psi^{\prime 2}\ket - \bra\psi^\prime\ket^2$ & $\sqrt{\bra\psi^{\prime 2}\ket
-\bra\psi^\prime\ket^2}$ \\
\hline
\vphantom{\Large A}%
$2\sqrt{2}$	& $1/2$	&	$0.013\pm 0.002$	&	$0.116\pm 0.009$ \\
$4$		& $1/2$	&	$0.014\pm 0.002$	&	$0.121\pm 0.010$ \\
$4\sqrt{2}$	& $1/2$ &	$0.016\pm 0.003$	&	$0.128\pm 0.012$ \\
$8$		& $1/2$	&	$0.018\pm 0.003$	&	$0.135\pm 0.013$ \\
\hline
\end{tabular}
\end{center}
\caption{The mean transparencies $\bra\psi^\prime\ket$, the variance $\bra\psi^{\prime 2}\ket - \bra\psi^\prime\ket^2$
and the standard deviation $\sqrt{\bra\psi^{\prime 2}\ket - \bra\psi^\prime\ket^2}$ together with their respective
uncertainties ($1\sigma$) for a set of GRFs encoding Gaussian PSFs with differing width $\sigma_x$.}
\label{table_psistat}
\end{table}

One would expect that with increasing PSF width $\sigma_x$ the variance decreases, which would be explained by the fact
that the variance is given by a weighted integration over the power spectrum $P(k)$. For increased position resolution,
i.e. a narrow PSF $c(\bmath{x})$, a wide power spectrum $P(k)$ is needed, which in turn would lead to a high variance.

This simple argument however, does not straightforwardly apply to the scaled realisations at hand: As laid down in
eqn.~(\ref{scaling2}), the field $\psi(\bmath{x})$ is modified by a factor depending on the maximal value
$\left|\psi(\bmath{x})\right|$ of the particular realisation. The occurence of a high amplitude is following a Gaussian
distribution with variance $\propto \int\dd^2 k\:P(k)$. This means, that in the case of narrow PSFs $c(x)$,
i.e. for wide power spectra $P(k)$, the field $\psi(\bmath{x})$ is more likely to assume large amplitudes 
\citep[compare][]{1956_peak_statistic}. The latter effect is of great importance and causes the surprising result that the 
measured variances in $\psi^\prime(\bmath{x})$ are larger for extended PSFs.

Comparing coded masks based on GRFs with purely random fields, the modulation of the
shadowgram decreases by a factor $\sim3\ldots4$. Therefore, the sensitivity is expected to be weaker. While the
above consideration is only valid for the observation of point sources, sensitivity is most likely to be gained in the
observation of extended sources. For those sources, it is possible to adjust the PSF to the expected source
intensity profile. In this case, modulations below the scale of the object to be observed are discarded - this
corresponds to applying Wiener filtering to the recorded shadowgram prior to source reconstruction.

\subsection{Distribution of the pixel amplitudes}\label{pixamp_distro}
\begin{figure}
\resizebox{\hsize}{!}{\includegraphics{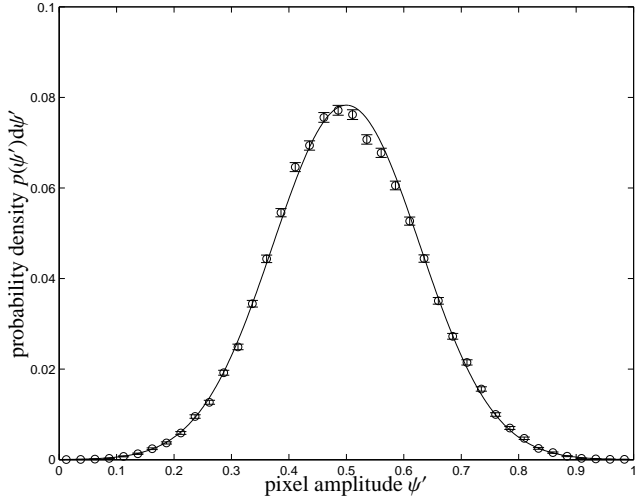}}
\caption{Probability density $p\left(\psi^\prime\right)\dd\psi^\prime$ of the pixel amplitudes
$\psi^\prime\left(\bmath{x}\right)$ (circles) and the best-fitting Gaussian for a particular realisation of a GRF. The error 
bars are Poissonian errors.}
\label{fig_psidistribution}
\end{figure}

As Fig.~\ref{fig_psidistribution} illustrates, the pixel amplitudes $\psi\left(\bmath{x}\right)$ follow a Gaussian
distribution, irrespective of the encoded PSF,
\begin{equation}
p\left(\psi^\prime\right)\dd\psi^\prime =
\frac{1}{\sqrt{2\pi}\sigma_\psi}
\exp\left(-\frac{\left(\psi^\prime-\mu_\psi\right)^2}{2\sigma^2_\psi}\right)\dd\psi^\prime\mbox{,}
\end{equation}
as a consequence of the central limit theorem (see \citet{central_limit_theorem}). The mean and variance of that
particular realisation have been determined to be $\mu_\psi = 0.5000\pm0.001$ and $\sigma_\psi=0.1277\pm0.0007$ at 95\%
confidence. For illustrative purposes, a Gaussian PSF with $\sigma_x=2\sqrt{2}\mbox{ pixels}$ has been chosen.

Again, it should be emphasised that the scaling eqn.~(\ref{scaling2}), while being reasonable from the physical point
of view, is not conserving Gaussianity. This is for the application at hand not a serious limitation, because the
variance of the distribution $p\left(\psi^\prime\right)\dd\psi^\prime$ is small compared to 1.

\subsection{Partial shadowing}\label{partial_shadowing}
It is interesting to see how partial shadowing affects shape and amplitude of the auto-correlation function. If a source
is observed at large off-axis angles, the shadowgram cast by the coded mask onto the detector is incomplete and
reconstruction artefacts emerge in the correlation function. In order to examine the extent to which the PSF suffers
from partial shadowing, the amplitudes $\psi_G(\bmath{x})$ in a margin amounting to a fraction of 25\%, 50\% and 75\% of
the total area have been set to zero and the cross-correlation function
$\psi_G(\bmath{x})\otimes\psi_G^{shadow}(\bmath{x})$ has been determined with the full coded mask.

\begin{figure}
\resizebox{\hsize}{!}{\includegraphics{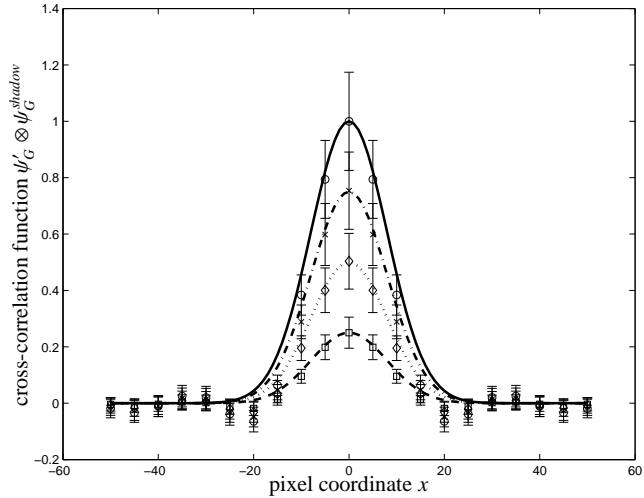}}
\caption{Cross-correlation function $\psi^\prime_G(\bmath{x})\otimes\psi_G^{shadow}(\bmath{x})$ and the respectively
expected PSF $c(x)$ with a shadowed margin corresponding to 25\% (dashed, squares), 50\% (dotted, diamonds) and 75\%
(dash-dotted, crosses) of the total area and, for comparison, the PSF for full imaging (solid, circles).}
\label{fig_partial_cross}
\end{figure}

As Fig.~\ref{fig_partial_cross} shows for a Gaussian PSF with $\sigma_x = 8\mbox{ pixels}$, the PSF drops in central
amplitude according the unshadowed area, but otherwise its shape remains unaltered. A second observation is that the
amplitude of the sidelobes is unaffected by the partial shadowing.

\begin{figure}
\begin{center}
\resizebox{7cm}{!}{\includegraphics{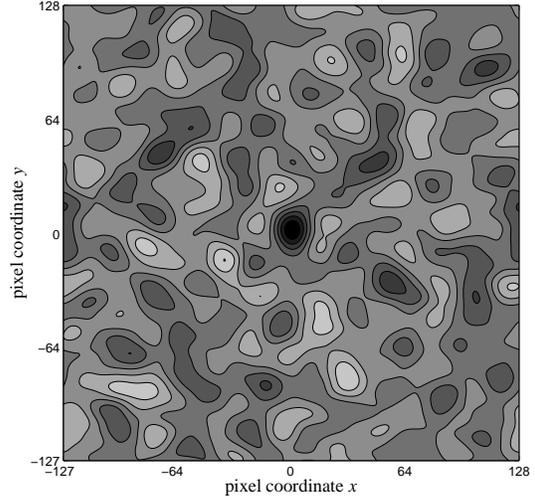}}
\end{center}
\caption{Reconstructed auto-correlation function $\psi^\prime_G(\bmath{x})\otimes\psi_G^{shadow}(\bmath{x})$ for a
shadowgram of which only $\simeq 3\mbox{\%}$ have been imaged onto the detector.}
\label{fig_partial_psf}
\end{figure}

The reconstructed PSF $\psi^\prime_G(\bmath{x})\otimes\psi_G^{shadow}(\bmath{x})$ for the case of radiation from a
source situated at large angles away from the optical axis, where only $1/32$ of the mask has been imaged onto the
detector is depicted in Fig.~\ref{fig_partial_psf}. Even though a tiny part of the mask amounting to $\simeq 3$\%
has been imaged, the correlation peak is clearly recognisable and its peak value is a factor $\simeq 4$ above the
correlation noise.

\subsection{Thresholded realisations}
\label{sec_threshold}
Due to possible technical complications in attempting to build a coded mask pattern based on a GRF with quasi-contiuous 
opaqueness, thresholded realisations are considered. A second argument in favour of thresholded realisations would be 
their achromatic properties, because the mask has to be constructed from the field $\psi^\prime(\bmath{x})$ for a specific
photon distribution in order to assure the maximal modulation of the shadowgram cast onto the detector. Yet another argument in 
favour of thresholded realisations of GRFs is their better sensitivity, because they imprint a stronger modulation of the 
shadowgram compared to smoothly varying GRFs.

\begin{figure}
\begin{center}
\begin{tabular}{c}
\resizebox{7cm}{!}{\includegraphics{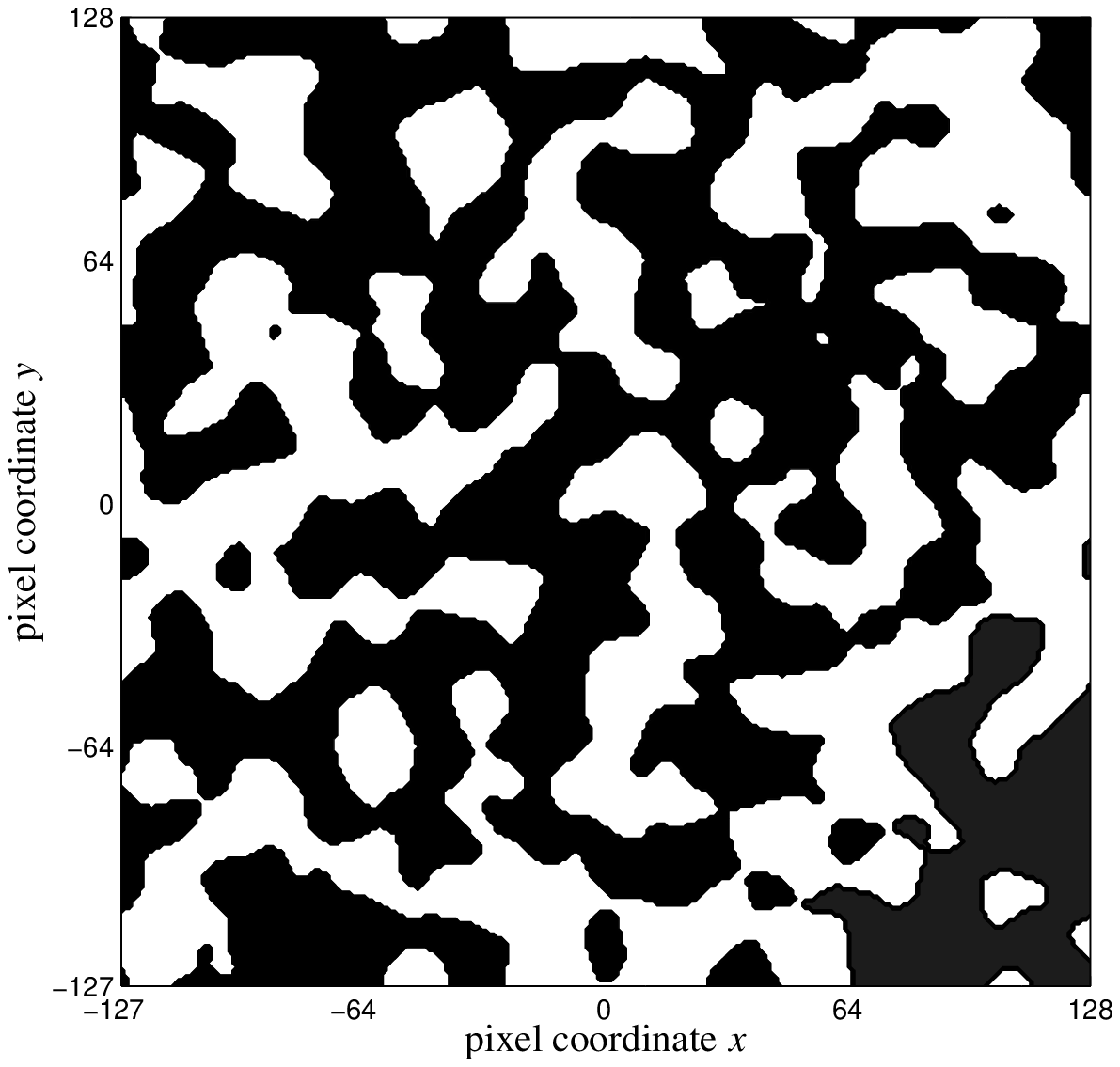}} \\
\resizebox{7cm}{!}{\includegraphics{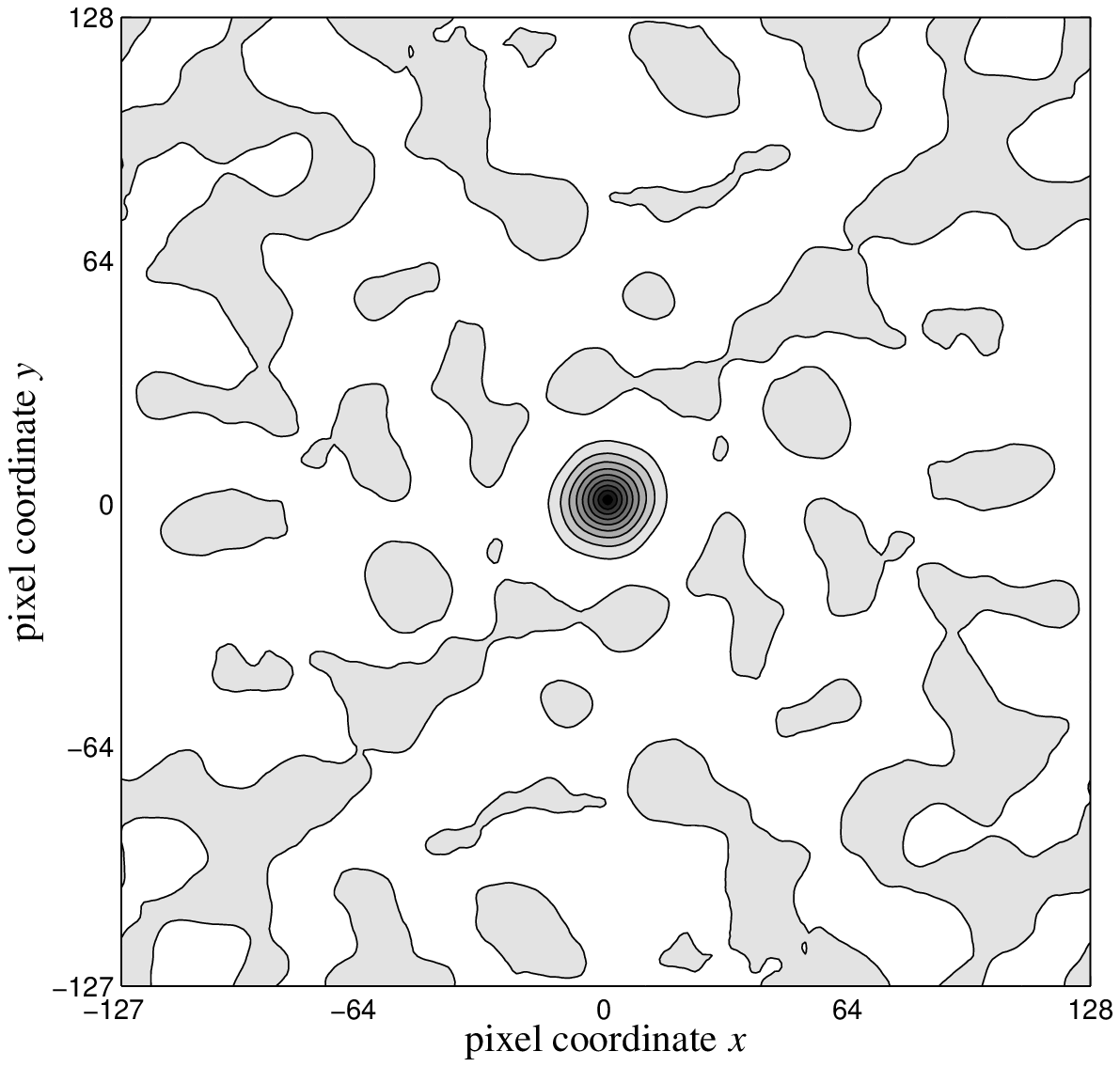}}
\end{tabular}
\end{center}
\caption{A thresholded realisation $\psi^{(t)}_G(\bmath{x})$ of a GRF $\psi_G(\bmath{x})$ (upper
panel) for the Gaussian PSF $c_G(x)$ and the auto-correlation 
function $\psi_G^{(t)}(\bmath{x})\otimes\psi_G^{(t)}(\bmath{x})$ (lower panel).}
\label{fig_digital_realise}
\end{figure}

In thresholded realisations, mask elements are taken to be transparent, if the value $\psi(\bmath{x})$ of the realisation is 
greater than zero, conversely, for values $\psi(\bmath{x})<0$ the mask element is set to be opaque. An example for a 
thresholded realisation of a GRF and its PSF is given in Fig.~\ref{fig_digital_realise}.

\begin{figure}
\resizebox{\hsize}{!}{\includegraphics{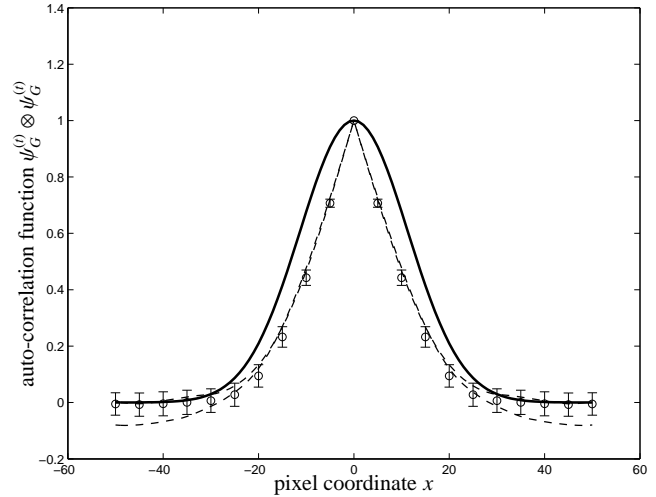}}
\caption{Deterioration of the auto-correlation function $\psi_G^{(t)}(\bmath{x})\otimes\psi_G^{(t)}(\bmath{x})$ from two
thresholded realisations $\psi_G^{(t)}(\bmath{x})$ (dashed) in comparison to the initial Gaussian PSF $c_G(x)$ (solid).}
\label{fig_digital_psf}
\end{figure}

An important issue is the degradation of the PSF $\psi_G^{(t)}(\bmath{x})\otimes\psi_G^{(t)}(\bmath{x})$ imposed by the
thresholding. As Fig.~\ref{fig_digital_psf} illustrates, the resulting auto-correlation function is pointy and its
kurtosis is larger than zero (leptokurtic). This results from the fact that small scale power is added by the
thresholding: In order to construct a step transition, more small-scale Fourier modes are needed, which leads to an
additive power law contribution $\propto k^{-2}$ in the power spectrum $P(k)$, such that the power spectrum acquires
Lorenzian wings. The point spread function $\psi_G^{(t)}(\bmath{x})\otimes\psi_G^{(t)}(\bmath{x})$, being the inverse
Fourier transform of $P(k)$, can then be approximated by two decaying branches of an exponential, which readily explains
the pointiness. The target Gaussian PSF $c_G(x)$ with $\sigma_x=8\sqrt{2}\mbox{ pixels}$ is shown for comparison. Again,
the error of the auto-correlation function is estimated by determining the sample variance in 100 realisations.

The size distribution of the patches as a function of threshold value can be described by means of the Press-Schechter
theory well known in cosmology. \citet{1974ApJ...187..425P}, \citet{1991ApJ...379..440B} and \citet{2002MNRAS.336..112M} 
provide the mathematical foundation.

\section{Ray-tracing simulations including finite photon statistics}\label{finite_photon}
Extensive ray-tracing simulations were performed describing the imaging of point sources with a finite number of photons 
(Sect.~\ref{finite_photon_statistics}), and the attainable sensitivity in such an observation was 
assessed (Sect.~\ref{ptsource_sensitivity}). The analogous was carried out for the observation of extended sources 
(Sect.~\ref{extsource_sensitivity}). Finally, the size of the field-of-view in the case of GRFs compared to 
traditional masks is examined (Sect.~\ref{decrease_fieldofview}). In the following, coded mask patterns based on Gaussian 
random fields are compared to purely random mask patterns and the mask pattern used in the WFI-instrument onboard {\em 
BeppoSAX}.

\subsection{Simulation setup}\label{finite_photon_statistics}
In the following, the performance of the coded mask is examined as a function of photon statistics. The statistical 
significance $\sigma$ of a simulated observation is defined to be
\begin{equation}
\sigma = \frac{N_\mathrm{source}}{\sqrt{N_\mathrm{bg}}}\mbox{,}
\end{equation}
where $N_\mathrm{source}$ and $N_\mathrm{bg}$ denote the source and background count rates, respectively. Here it 
should be emphasised, that $\sigma$, $N_\mathrm{source}$ and $N_\mathrm{bg}$ always refer to the number of actually 
detected photons which makes a difference when considering the coded mask employed in {BeppoSAX}'s WFI instrument, in which the 
average transparency is not equal to $1/2$.

Observations were simulated by randomly choosing $2 N_\mathrm{source}$ homogeneously distributed photon impact positions 
$\bmath{x}$ across the mask face. In order to emulate the random process of photons penetrating the mask, a 
homogeneously distributed random number $r$ from the interval $r\in [0\ldots1]$ was drawn for each photon, and compared to the 
value $\psi(\bmath{x})$ of the GRF at the same position $\bmath{x}$. In the case $r>\psi(\bmath{x})$ the photon was assumed to 
be able to penetrate the mask, whereas in the case $r\leq\psi(\bmath{x})$ the photon was taken to be absorbed by the mask. For 
{\em BeppoSAX}'s mask pattern, which has an average transparency of $1/3$, a total number of $3 N_\mathrm{source}$ photons was 
simulated. 

For the background, which was assumed to be homogeneous, $N_\mathrm{bg}$ photon impact positions were determined and 
the count rates in the corresponding pixels were increased accordingly. Background count rates were fixed to a value of 
$N_\mathrm{bg}=10^4$~photons, which are typical for an instrument like WFI in a 100~second exposure.

The resulting field $\psi^{(sim)}$ containing the number of photons that struck a certain pixel was then correlated with the 
original mask pattern $\psi$, again using balanced correlation. In the next step, the highest peak was localised in the 
correlated data field $\psi^{(sim)}\otimes\psi$ and its significance $\Sigma$ was determined by comparing the peak height 
$a_\mathrm{max}=\mathrm{max}\{\psi^{(sim)}\otimes\psi\}$ to the level of fluctuations  
$\sigma_\psi^2=\bra\left(\psi^{(sim)}\otimes\psi\right)^2\ket$ in the field. If the peak had a significance 
$\Sigma=a_\mathrm{max}/\sigma_\psi$ exceeding $3$ and was located at a position which deviated less than half a PSF 
width from the nominal position, the simulated detection was taken to be successful. A particular realisation of a Gaussian 
random field was exposed to 100 simulated photon distributions from which the detection probability $p$ (i.e. the occurence of 
a $\geq3\sigma_\psi$-peak located at the correct position) and the false detection probability $q$ (i.e. the occurence of a 
$\geq3\sigma_\psi$-peak at a wrong position) was derived. The sample variance in comparing 100 realisations of Gaussian random 
fields was used to derive errors on $p$ and $q$. For the purpose of this work, the detector efficiency and position response 
were assumed to be ideal.

\subsection{Point source sensitivity of a set of Gaussian random fields}\label{ptsource_sensitivity}
\begin{figure}
\resizebox{\hsize}{!}{\includegraphics[angle=270]{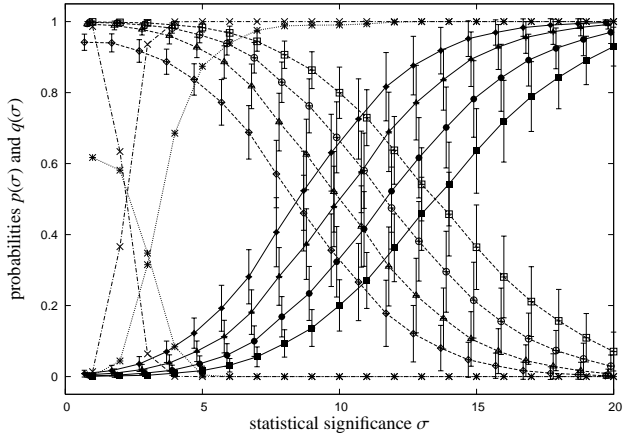}}
\caption{Point source sensitivity in on-axis observations of a set of GRFs: The detection probability 
$p(\sigma)$ (solid lines, closed symbols) and the false detection probability $q(\sigma)$ (dashed lines, open symbols) are 
plotted as functions of statistical significance $\sigma$ for PSF widths $\sigma_x=2$ (squares), $\sigma_x=2\sqrt{2}$ 
(circles), $\sigma_x=4$ (triangles) and $\sigma_x=4\sqrt{2}$ (diamonds), in comparison to purely random masks (dotted line, 
stars) and {\em BeppoSAX}-WFI pattern (dash-dotted line, crosses). In contrast to the ensemble of GRFs it is 
not possible to state an ensemble variance for $p(\sigma)$ and $q(\sigma)$ in the case of {\em BeppoSAX}'s pattern. The data 
points have been slightly displaced for better visibility.}
\label{fig_pt_finite_photon}
\end{figure}

Fig.~\ref{fig_pt_finite_photon} shows the detection probability $p$ and the false detection probability $q$ as a function of 
photon statistics, expressed in terms of statistical significance $\sigma$ for GRFs, a purely random mask and 
{\em BeppoSAX}'s URA pattern. The source was assumed to lie on the optical axis, i.e. the mask pattern is imaged completely 
onto the detector. Common to all mask patterns is the fact that $p$ rises with statistical significance, and that $q$ drops 
accordingly. But while reliable observations can be done using the {\em BeppoSAX}-pattern or random patterns even at low 
photon statistics of $2-3\sigma$, the patterns based on GRFs require high photon fluxes. For them, 
observations are feasible starting from $\simeq9\sigma$. The reason why GRFs are less sensitive to the 
traditional mask pattern is the fact that they imprint a weaker modulation of the shadowgram. Furthermore, one immediately 
notices the trend that the patterns are more sensitive for wider PSF widths due to the increase in variance of the mask 
pattern with increasing PSF width. Thus, position resolution is traded for sensitivity.

\begin{figure}
\resizebox{\hsize}{!}{\includegraphics[angle=270]{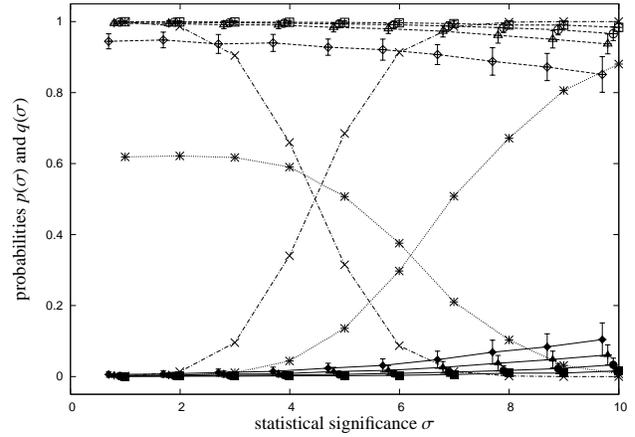}}
\caption{Point source sensitivity in off-axis observations (50\% obscuration) of a set of GRFs: The detection 
probability $p(\sigma)$ (solid lines, closed symbols) and the false detection probability $q(\sigma)$ (dashed lines, open 
symbols) are given for PSF widths $\sigma_x=2$ (squares), $\sigma_x=2\sqrt{2}$ (circles), $\sigma_x=4$ (triangles) and 
$\sigma_x=4\sqrt{2}$ (diamonds), in comparison to purely random masks (dotted line, stars) and {\em BeppoSAX}'s WFI pattern 
(dash-dotted line, crosses).}
\label{fig_pt_obscure}
\end{figure}

Fig.~\ref{fig_pt_obscure} shows the analogous results for an off-axis observation in which only half of the mask pattern has 
been imaged onto the detector. The result corresponds to the findings for the case of normal incidence, but $p(\sigma)$ and 
$q(\sigma)$ are shifted to higher values of $\sigma$, which is due to the fact, that only half of the photons actually reach 
the detector and that the reconstruction has to cope with the decreased signal. Again, one attains higher sensitivities for 
wider PSFs in the case of patterns based on GRFs.

Common to all figures is the fact, that the curves $p(\sigma)$ and $q(\sigma)$ are not adding up to one, which is caused by the 
combined criterion where apart from the correct peak position a minimal peak height above the correlation background is 
required, which is often not fulfilled in the cases of low photon statistics.

\subsection{Sensitivity in observations of extended sources}\label{extsource_sensitivity}
In addition, suitable simulations were carried out in order to assess the performance of GRFs in the 
observation of extended sources, such as supernova remnants, structures in the Milky Way and clusters of galaxies. Typical 
sizes of those sources range between arcminutes and a degree. For simplicity, the source was assumed to be described by a 
Gaussian profile with extension $\sigma_\mathrm{profile}=2$~pixels. The shadowgram recorded in observations of extended sources 
are superpositions of slightly displaced point source shadowgrams, where the relative intensities follow from the source 
profile. Consequently, the imaging of extended sources is simulated by convolving the mask pattern with the source profile 
prior to the ray-tracing. Despite that, the image reconstruction has been carried out with the unconvolved mask pattern.

\begin{figure}
\resizebox{\hsize}{!}{\includegraphics[angle=270]{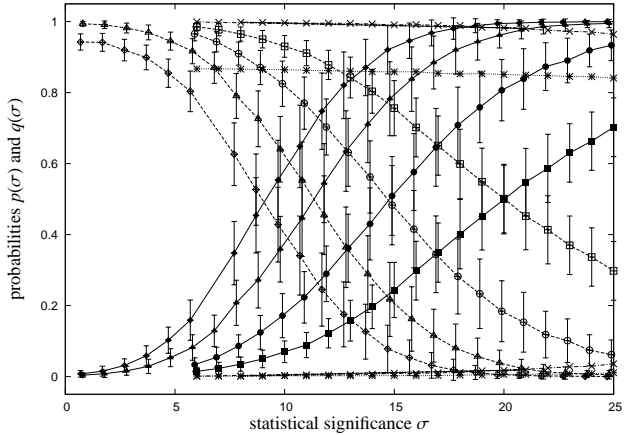}}
\caption{Sensitivity in on-axis observations of extended sources of a set of GRFs: The detection probability 
$p(\sigma)$ (solid lines, closed symbols) are given along the false detection probability $q(\sigma)$ (dashed lines, open 
symbols) for PSF widths of $\sigma_x=2$ (squares), $\sigma_x=2\sqrt{2}$ (circles), $\sigma_x=4$ (triangles) and 
$\sigma_x=4\sqrt{2}$ (diamonds), in comparison to purely random masks (dotted line, stars) and {\em BeppoSAX}'s WFI pattern 
(dash-dotted line, crosses).}
\label{fig_ext_sigma}
\end{figure}

Fig.~\ref{fig_ext_sigma} gives the dependence of the detection probability $p$ and the corresponding $q$ on the photon counting 
statistic $\sigma$. In the observation of extended sources, the patterns based on GRFs are superior to the 
traditional approaches: While reliable detections can be achieved starting from $\sigma\gsim10$ (for $\sigma_x=4\sqrt{2}$) up 
to $\sigma\gsim20$ (for $\sigma_x=2$), the performance of the traditional masks is notably worse. At the examined levels of 
photons statistics, the detection probability $p$ stays close to zero and shows but a shallow increase with $\sigma$ in the 
case of {\em BeppoSAX}'s URA pattern. 

The good performance of the GRFs, and their decreasing performance with correlation length, i.e. PSF width 
$\sigma_x$ is of course to be traced back to the fact, that mask patterns with large structures are less affected by the 
convolution with the source profile than mask patterns exhibiting small structures; in the extreme case of random masks or 
{\em BeppoSAX}'s pattern, the structures are washed out and consequently, the modulation of the shadowgram is very weak. This 
can be circumvent, however, by tuning the angular size of a mask pixel to match the angular size of the source to be observed.

\subsection{Field-of-view in the observation of point sources}\label{decrease_fieldofview}
Now, the size of the field-of-view, i.e. the minimal fraction of the mask pattern required to be imaged onto the 
detector in order to yield a significant detection peak is investigated. For that purpose, the point source detection 
probability $p$ and the false detection probability $q$ are considered to be functions of the obscuration $Q$, which is defined 
as the fraction of the mask area imaged onto the detector. The number of background photons was kept fixed to be 
$N_\mathrm{bg}=10^4$, while the number of source photons $N_\mathrm{source}$ was diminished by this factor of $Q$ prior to the 
ray-tracing. Their number was fixed to yield a significance of $\sigma=20$ for $Q=1$, i.e. for the case of complete imaging. 
The background photons were assumed to be homogeneously distributed. The simulation and the derivation of the values for $p(Q)$ 
and $q(Q)$ were carried out in complete analogy to Sect.~\ref{ptsource_sensitivity}.

\begin{figure}
\resizebox{\hsize}{!}{\includegraphics[angle=270]{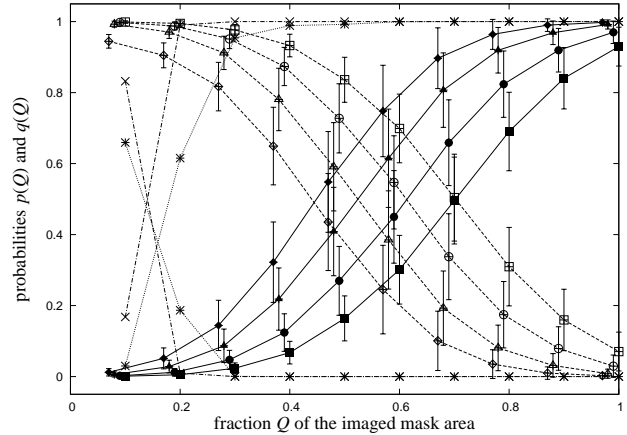}}
\caption{Decrease of the field-of-view: The source detection probability $p(Q)$ (solid lines, closed symbols) and the false 
detection probability $q(Q)$ (dashed lines, open symbols) as functions of the obscuration $Q$ are shown for a set of Gaussian 
random fields for varying PSF width: $\sigma_x=2$ (squares), $\sigma_x=2\sqrt{2}$ (circles), $\sigma_x=4$ (triangles) and 
$\sigma_x=4\sqrt{2}$ (diamonds). In comparison, a purely random mask (dotted line, stars) and {\em BeppoSAX}'s WFI pattern 
(dash-dotted line, crosses) are considered.}
\label{fig_fof_decrease}
\end{figure}

The results are depicted in Fig.~\ref{fig_fof_decrease}: While the traditional patterns show a good performance and have a high 
detection probability $p(Q)$ for values of $Q\gsim0.1$ ({\em BeppoSAX}'s pattern) and $Q\gsim0.2$ (random mask), the 
GRFs fall behind significantly in performance. Imaging is only possible in the cases where a fraction of at least 
$Q=0.5\ldots0.6$ of the mask has been imaged onto the detector, resulting in a decrease of the field-of-view of about a factor 
of $3\ldots5$, which renders the usage of GRFs very unlikely in survey missions. Again, the GRF patterns encoding wide PSFs are 
more sensitive and yield larger fields-of-view than GRFs with narrow PSFs.

\section{Summary and outlook}\label{summary}
In this article, a new algorithm for generating coded masks is presented that allows the construction of a mask with
defined imaging properties, i.e. point spread functions.

\begin{itemize}
\item{The viability of constructing a coded mask for a predefined PSF as a realisation of a GRF has
been shown. For realisations generated with differing random seeds, the shape of the PSF is reproducible with high
accuracy. Due to the reproducibility of the PSF, the parameter space is greatly reduced and the necessity of running
extensive Monte-Carlo simulations is alleviated.}

\item{The generation of 2-dimensional URA patterns requires the number of pixels in each direction to be
incommensurable, i.e. they are not allowed to have a common divisor. While twin prime numbers exist, mask patterns
generated that way are almost, but not quite square \citep{1977SSI.....3..473M,1979MNRAS.187..633P}. Coded masks based on 
GRFs may have any side length and any ratio of side lengths. Additionally, sizes chosen equal to $2^n$, 
$n\in\mathbb{N}$ enable the usage of fast Fourier transforms. Realisations of GRFs have cyclic boundary 
conditions which is a desirable feature for coded mask imagers.}

\item{The average transparency of coded mask patterns based on scaled GRFs is equal to 1/2, irrespective of the PSF 
they encode. The pixel amplitudes of a realisation are Gaussianly distributed as a consequence of the central limit theorem. 
The pixel-to-pixel variance, however, is smaller in the case of GRFs compared to purely random fields, which 
results in a weaker modulation of the shadowgram and hence the sensitivity is expected to be smaller. The variance shows the 
trend of decreasing with increasing PSF width, which is caused by the scaling with the maximal values of the realisation.}

\item{Coded masks based on GRFs are chromatic in contrast to purely random fields: The mask pattern
has to be designed for a specific spectral distribution of photons due to semi-transparent mask elements. Any mismatch
in the photon spectrum of a source under observation would result in a less pronounced modulation of the shadowgram,
which in turn affects the sensitivity of the coded mask imager. A possible workaround is the usage of thresholded Gaussian 
random fields, that show pointy auto-correlation functions in contrast to smooth target PSFs. Another advantage is their 
enhanced sensitivity due to the stronger modulation of the shadowgram. The properties of thresholded realisations, however, 
show a large sample variance which requires selections with suitable criteria after construction.}

\item{Ray-tracing simulations including finite photons statistics and background noise show, that the sensitivity of 
GRFs falls behind that of purely random masks and URA patterns like the one employed in {\em BeppoSAX} by a 
factor of $2\ldots3$ in the observation of {\em point sources}, depending on PSF width. For GRFs, the sensitivity was found to 
depend exponentially on PSF width, one is trading sensitivity for position resolution.}

\item{The sensitivity of patterns based on GRFs is significantly better in the observation of {\em extended sources} because 
their comparably large structures are less affected by the convolution with the source profile than traditional masks that 
possess pronounced structures on small scales.}

\item{Finally, the size of the field-of-view of GRFs in comparison to traditional masks is examined. It is 
found that reliable imaging can only performed with GRFs, if a large fraction of the mask is imaged onto the 
detector.  In contrast, purely random masks and especially {\em BeppoSAX}'s URA pattern enable imaging at large off-axis 
angles. Comparing the resulting fields-of-view for the preset number of photons shows, that the field-of-view of patterns based 
on GRFs are smaller by a factor of $3\ldots5$ (depending on PSF width).}

\end{itemize}

Although the shortcomings of Gaussian random fields with respect to point source sensitivity, chromaticity and localisation 
accuracy make their usage in observing point sources doubtful, they may find application in observations of 
extended sources, while simultaneously providing a moderate performance in the observation of point sources. Coded mask 
patterns on the basis of GRFs may be aesthetically pleasing because they utilise an abstract cosmological concept for a 
technological application.

\section*{Acknowledgments}
I would like to thank Matthias Bartelmann and Christoph Pfrommer for careful reading of the manuscript, Eugene
Churazov for constructive and valuable comments and Saleem Zaroubi for many clarifying discussions. Furthermore, I am 
grateful to Kazuo Makishima for giving me details of the WFI experiment on {\em BeppoSAX}.

\bibliography{bibtex/aamnem,bibtex/references}
\bibliographystyle{mn2e}

\appendix

\bsp

\label{lastpage}

\end{document}